\documentclass[a4paper,11pt]{article}

\usepackage{pos_MST}
\usepackage{lipsum} 

\usepackage{subcaption}
\usepackage{graphicx}
\usepackage{booktabs}
\usepackage{multirow}

\title{Status of the Medium-Sized Telescopes for the Cherenkov Telescope Array Observatory}

\ShortTitle{Status of the Medium-Sized Telescopes for CTAO}

\author*[1]{Federica Bradascio}

\affiliation[1]{IRFU, CEA, Universit\'e Paris-Saclay, F-91191 Gif-sur-Yvette, France}

\forColl{CTA MST} 
\emailAdd{federica.bradascio@cea.fr}

\abstract{

The Cherenkov Telescope Array Observatory (CTAO) is a next-generation ground-based gamma-ray observatory that will study the universe at very high energy using atmospheric Cherenkov light. CTAO will comprise over 67 telescopes of three different sizes, located in the northern and southern hemispheres. Among these, the Medium-Sized Telescope (MST) will play a crucial role in CTAO's observations, providing excellent sensitivity and angular resolution for gamma rays in the energy range of 100 GeV to 5 TeV. The MST is based on a modified single-mirror Davies-Cotton design, featuring a segmented mirror with a diameter of 12 meters, a total reflective surface of 88 square meters, and a focal length of 16 meters. It will cover an approximately 8-degree field of view and be equipped with two different cameras: NectarCAM and FlashCam, at the northern and southern CTAO sites, respectively. The MST's design is optimized for efficient observation of extended sources, including supernova remnants and pulsar wind nebulae, as well as the study of gamma-ray bursts and active galactic nuclei. 
{Currently, the MST is in the midst of production and testing stages with the aim of being installed in 2025 for the CTAO Pathfinder project. In this project, one MST telescope will be deployed at each CTAO site to provide on-site experience with pre-production components. This approach facilitates cost and risk reduction before starting serial production.} This contribution provides an overview of the current status and plans of the MST's construction at both the northern and southern CTAO sites, as well as details on the telescope and camera designs and their expected performance. 

\ConferenceLogo{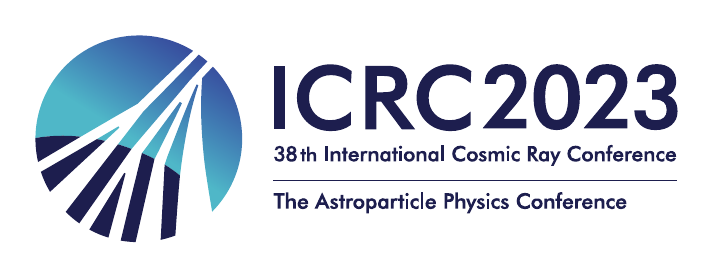}

\FullConference{The 38th International Cosmic Ray Conference (ICRC2023)\\ 26 July -- 3 August, 2023\\ Nagoya, Japan}
}

\begin{document}

\maketitle

\section{Introduction}

The Cherenkov Telescope Array Observatory (CTAO) ~\cite{CTAconcept, CTAdesign} is the next-generation ground-based observatory dedicated to gamma-ray astronomy across an energy range of 30~GeV to 300~TeV. It comprises two observatories: CTA South (CTAS) in Chile and CTA North (CTAN) in Spain, each employing telescopes of different sizes and designs to ensure comprehensive coverage of the sky, including galactic and extra-galactic sources. To achieve cost-effectiveness and wide energy coverage, the CTAO utilizes a combination of three telescope types. 
The Medium-Sized Telescopes (MSTs), distributed between both observatories cover the core energy range of 100~GeV to 5~TeV. The energy range below 100~GeV is extended by Large-Sized Telescopes (LSTs), while Small-Sized Telescopes (SSTs) cover the energy range above a few TeV.

This contribution focuses on providing an overview of the MST telescope's current state. MST is designed with specific characteristics, including a field of view (FoV) greater than 7 degrees, an optical point spread function (PSF) smaller than the pixel size of the Cherenkov camera (0.18 degrees), an effective mirror area larger than 88~m$^2$, a sources localization precision of 7 arcseconds, and the capability to point to any coordinates on the sky within 90 
seconds.  The MST’s design is optimized for efficient observation of extended sources, including supernova remnants and pulsar wind nebulae, as well as the study
of gamma-ray bursts and active galactic nuclei.

\begin{figure}
     \centering
     \includegraphics[width=\textwidth]{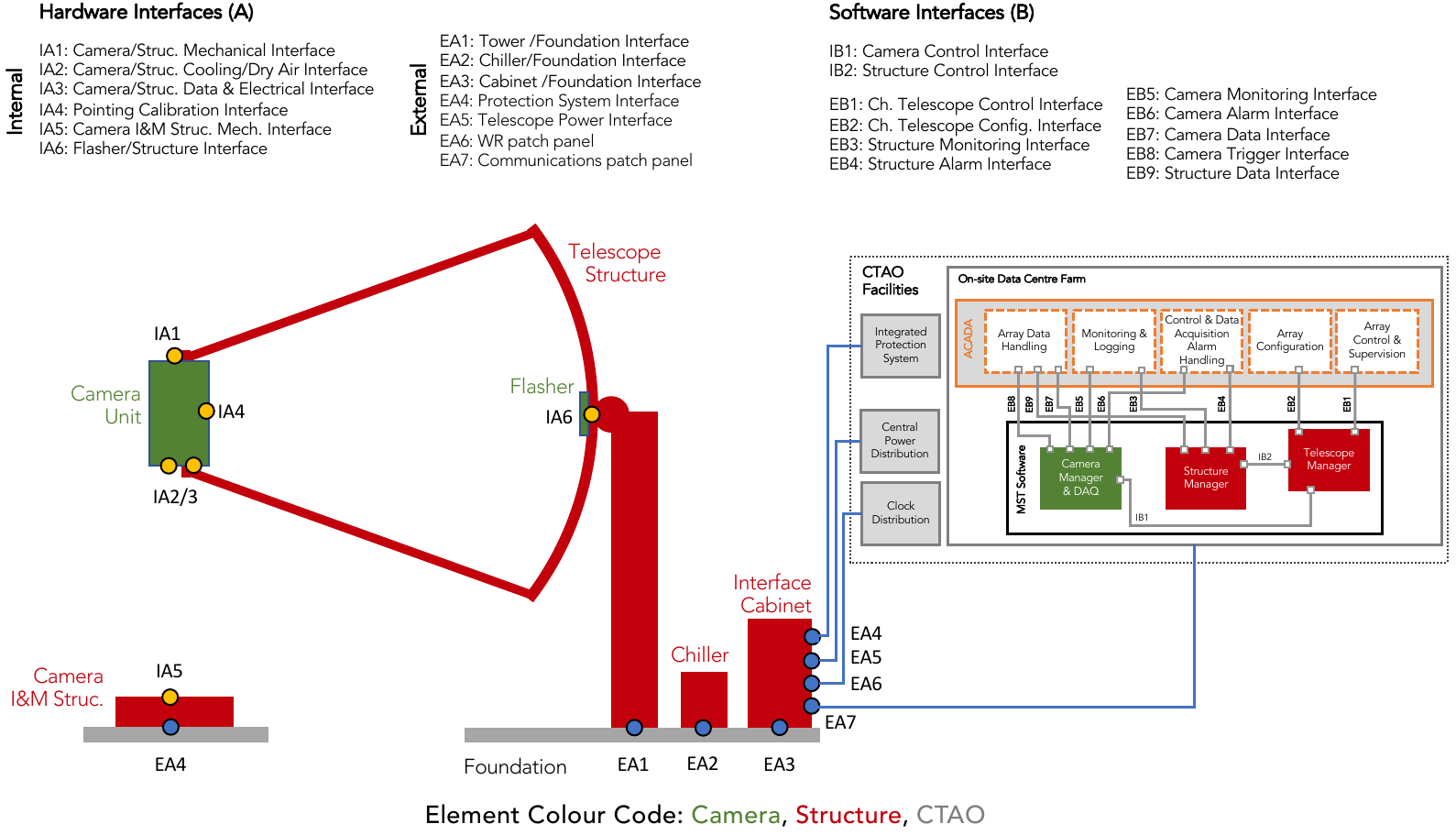}
     \caption{Overview of the MST design. It consists of three main components, each represented by a different color: the camera (green), the structure (red), and the infrastructure (gray). Additionally, a comprehensive list of all hardware and software interfaces is provided.}
     \label{fig:infrastructure}
 \end{figure}

The design of the MST, including its various components, is depicted in \autoref{fig:infrastructure}. MST comprises three main components, each distinguished by a different color: the camera (green), the structure (red), and the infrastructure (gray). In the following sections, we will provide a detailed status update for each of these components.

\section{Telescope structure}
\begin{figure}
\begin{center}
     \begin{subfigure}[c]{0.5\textwidth}
         \centering         \includegraphics[width=0.9\textwidth]{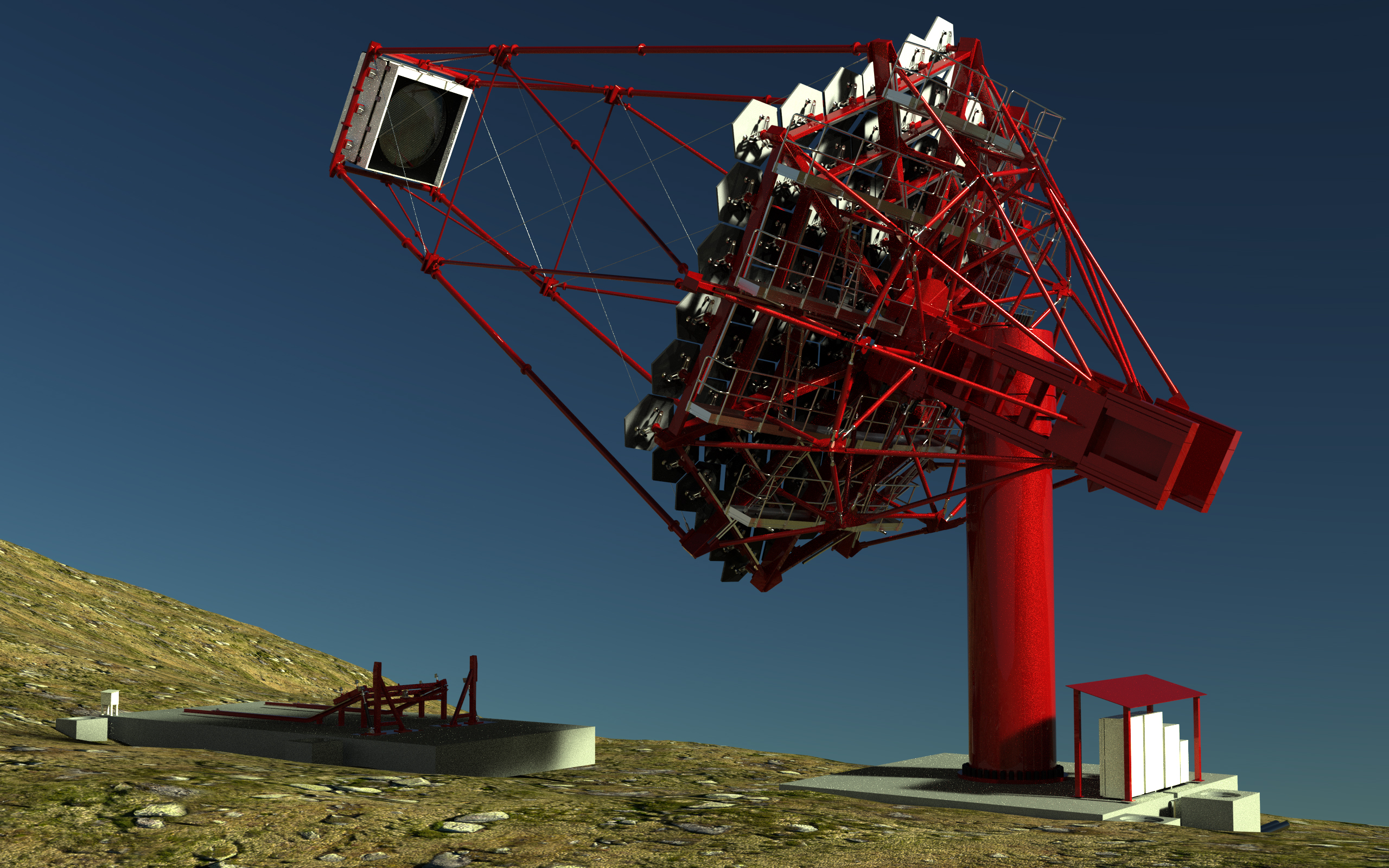}
        \caption{}
        \label{fig:mst_pic}
     \end{subfigure}%
     \hfill
    \hfill
     \begin{subfigure}[c]{0.5\textwidth}
         \centering
         \includegraphics[width=0.95\textwidth]{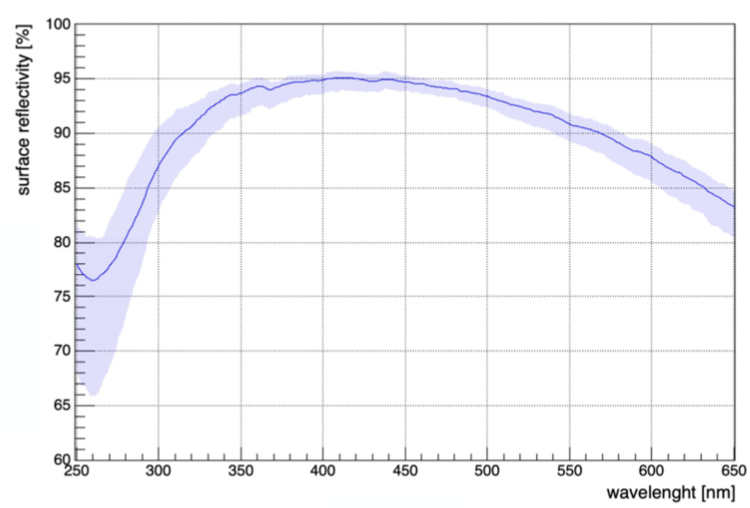}
        \caption{}
        \label{fig:mirrors_reflec}
     \end{subfigure}%
    \caption{\textbf{(a)}: Simulated picture of the MST structure. \textbf{(b)}: Mirrors' reflectivity as a function of wavelenght. The measurements were made on IRFU-Kerdry mirrors.}
    \label{fig:mst}
    \end{center}
\end{figure}

MST utilizes a modified single mirror Davies-Cotton (DC) layout~\cite{DAVIES195716}.  This design choice represents a tradeoff aimed at optimizing the PSF across a significant portion of the Cherenkov camera's FoV and improving the uniformity of the reflector's performance (i.e. isochronicity). \autoref{fig:mst_pic} illustrates a simulated depiction of the MST structure. The first MST structure was constructed in Berlin-Adlershof in 2013. Currently, the components of the MST structures for the Pathfinder on both CTAS and CTAN sites are in production.

The telescope structure is constructed using steel to ensure optimal rigidity of the optical support structure, thereby eliminating the need for frequent realignment of the mirrors to compensate for structural deformations during observations. One of its key components is the telescope's positioner (see element EA1 in \autoref{fig:infrastructure}), which facilitates the pointing and tracking of celestial objects in the sky. This positioner  is designed as a cylindrical tower, measuring 2 meters in diameter and 9 meters in height. It serves as a platform for the majority of electrical cabinets and is divided into three floors.
To enable smooth rotation of the optical support structure around the azimuth axis, a large bearing adapted from wind turbines acts as the interface between the tower and the head. This bearing facilitates the necessary azimuthal movement during observations. Additionally, the camera support structure's tubes are designed to possess sufficient stiffness, thereby minimizing both telescope flexion and shadowing of the mirrors.

The optical system consists of a tesselated reflector, comprising 86 hexagonally-shaped mirrors with a flat-to-flat side length of 1.2~m. The telescope has a focal length of 16~m. Each mirror has a nominal radius of curvature of 32.14 m, approximately twice the focal length. The reflector's larger radius of curvature of 19.2~m compared to the focal distance enables the focusing of light over 80\% of the FoV with a root mean square (RMS) optical time spread $<0.8$~ns. The mirrors are aligned to reflect rays parallel to the optical axis into the focal point. The mirror shape is achieved by bending a thin glass sheet onto a precisely formed mold, a technique called ``cold slumping''. The smoothness of the surface, and consequently the quality of the focused spot, is determined by the thickness and quality of the glass sheet, as well as the precision of the mold shape. To reinforce the mirror, an aluminum honeycomb sandwich panel is glued to the glass, with a second glass piece added on the other side of the honeycomb. The radius of curvature achieved through the cold slumping technology closely matches the nominal value, deviating only by a few centimeters. In order to withstand harsh environmental conditions such as rain and sand abrasion, the mirror facets are coated with a protective multilayer consisting of SiO$_2$ and either HfO$_2$ or ZrO$_2$. The mirror reflectivity exceeds 85\% within the range of 300 nm to 550 nm (see \autoref{fig:mirrors_reflec}), meeting the CTAO requirements. The mirrors are produced in France, Italy and Poland~\cite{mirror1, mirror2}.

The MST structure includes also the pointing calibration, which uses a single, wide FoV CCD camera (SingleCCD) installed in the center of the dish, aligned to the optical axis of the telescope and facing the Cherenkov camera. The FoV of this camera is chosen sufficiently large to observe both the star field (to determine the direction of the optical axis of the SingleCCD) and the pointing LEDs mounted on the Cherenkov camera body (to determine its orientation w.r.t. the optical axis).


\section{Cameras}
The MST telescopes in the two CTAO sites will be equipped with two distinct cameras: FlashCam for CTAS and NectarCAM for CTAN. These cameras must meet the following stringent requirements: event timing for large signals with a precision better than 2 ns, recording of the full waveform of triggered events within 60 ns, dynamic range of each pixel from 0 to 2000 photoelectrons (p.e), instrument deadtime fraction of less than 5\%, and gain calibration precision of less than 8\%. Meeting these specific requirements is crucial for ensuring optimal performance and data quality of the cameras in the MST telescopes.

\subsection{FlashCam}
FlashCam is based on a fully-digital design of the readout and trigger system~\cite{flashcam_WERNER201731}. It utilizes 1758 active pixels, with vacuum photo-multiplier tubes (PMTs) as photodetectors arranged into 147 modules, constituting the photon detection plane (PDP). These PDP modules provide high voltage for the photomultipliers, along with a pre-amplifier and an interface for slow control, monitoring, and safety functions. Winston cones are installed in front of each PMT to improve photon collection efficiency and restrict the sensitive angular range. A UV-transparent window is also installed in front of the PDP to protect the PMTs and Winston cones. To ensure stable operation and minimize gain changes, a water cooling system maintains the median PDP temperature at around 30$^{\circ}$~C, protecting the electronics.

The PDP module preamplifier operates in its linear transfer regime up to approximately 250~p.e., after which it enters a controlled saturation regime for higher amplitudes. In the saturation regime, the input charge is reconstructed from the area of the amplifier output. Integrating and unfolding these signals enables a single signal path for a large amplitude range, reducing components, costs, and power consumption. This design choice allows for an extended dynamic range beyond 3000~p.e. while maintaining linearity and sub-p.e. resolution for small signals~\cite{flashcamICRC2021}. 

FlashCam optimizes bandwidth requirements by utilizing only one gain channel. The signals are digitized with 2050 MSample/sec FADCs with a 12-bit dynamic range. The trigger evaluation is performed directly on the data, while trigger creation and additional pre-processing of digitized signals are handled by low-cost commercial FPGAs in the front-end electronics. During triggering, waveforms of time slices up to 128~ns in length are transmitted to a camera server in normal operations via four 10-Gbit Ethernet fibers, while two additional fibers are dedicated to slow control monitoring and clock distribution. For ease of installation and maintenance, the camera features rear doors that provide access to the racks. The photon detector plane modules can be installed from within the camera, eliminating the need to remove the optical front system for maintenance.

\begin{figure}
\begin{center}
     \begin{subfigure}[c]{0.5\textwidth}
         \centering
         \includegraphics[width=0.9\textwidth]{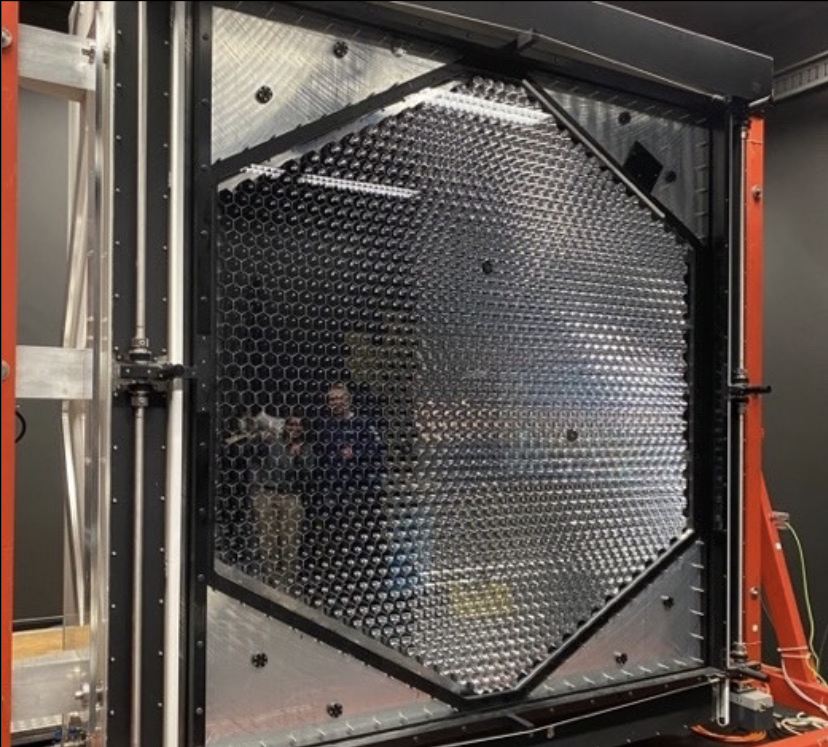}
        \caption{}
        \label{fig:flashcam_pic}
     \end{subfigure}%
     \hfill
     \begin{subfigure}[]{0.5\textwidth}
         \centering
         \includegraphics[width=0.95\textwidth]{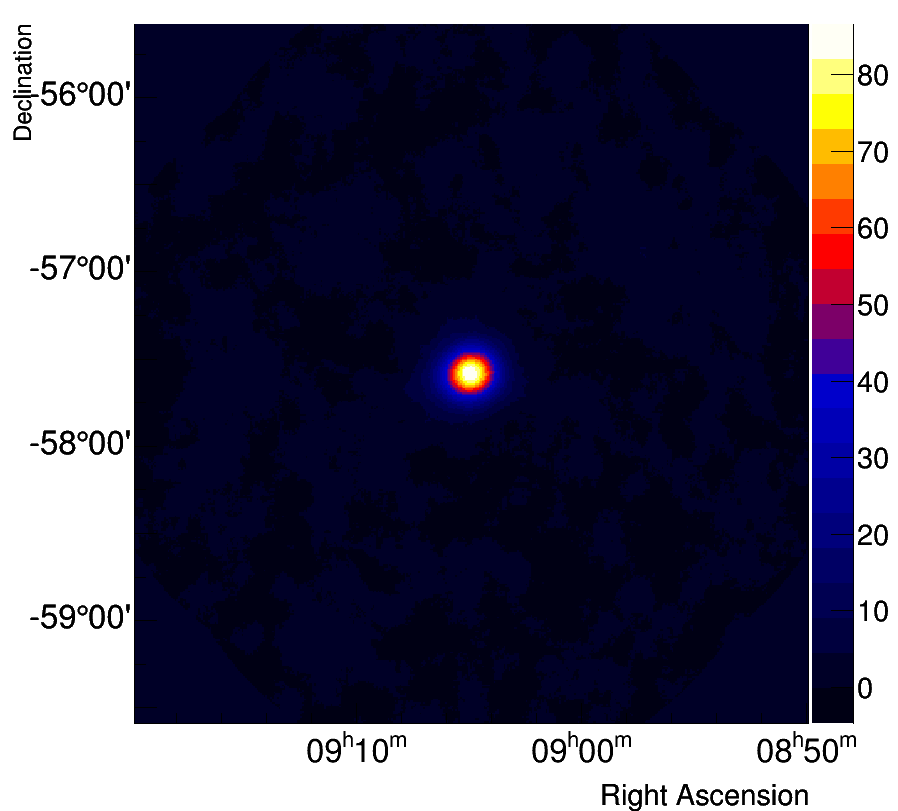}
        \caption{}
        \label{fig:flashcam_plot}
     \end{subfigure}%
    \caption{\textbf{(a)}: Current image of FlashCam at MPIK. \textbf{(b)}: Significance map of the field of view of PKS 0903-57 observed with CT5-FlashCam.}
    \label{fig:flashcam}
    \end{center}
\end{figure}

An advanced version of the prototype has been successfully installed on H.E.S.S.-CT5 in October 2019, regularly collecting physics data with an uptime of approximately 98\% since then~\cite{hess5_scienceprogram}. 
After commissioning FlashCam in CT5, a dedicated observation program was conducted to validate its performance by verifying simulations against actual data, checking pointing corrections, comparing measured point spread functions to simulations, examining background distribution, reconstructing energy spectra, and verifying event time stamping. In \autoref{fig:flashcam_plot}, the significance map obtained from the observations of PKS 0903-57 is depicted. These observations were triggered by an outburst of the source detected with \textit{Fermi}-LAT, leading to the discovery of a strong TeV gamma-ray flux from the source. The significance sky map was computed on a binwise basis using an oversampling circle with a radius of 0.1$^\circ$ for on-source events, while background events were collected in rings around each bin position, with an inner radius of 0.5$^\circ$ and an outer radius of 0.6$^\circ$~\cite{hess5_scienceprogram}. The Pathfinder camera for CTAS shown in \autoref{fig:flashcam_pic} is being finalized, awaiting the delivery of the final PMT modules. Once ready, the camera in its current configuration can be shipped to the Pathfinder site in 2024.

\subsection{NectarCAM}
The NectarCAM camera has a modular design, with a basic element (``module") consisting of a focal plane module (FPM) and a front-end board (FEB). The focal plane module is composed of seven 1.5" R12992-100-05 Hamamatsu photomultiplier tubes (PMTs) associated with high voltage and pre-amplification boards (HVPA), and equipped with Winston cone light concentrators. The Cherenkov light impinging on the camera is first detected in the focal plane, converted into an electric signal by the PMTs and preamplified towards two gain channels. The signal is then amplified a second time in the FEB by a custom amplifier called ACTA, where it is split into three channels: low and high gain channels, and trigger channel (level 0 or L0 trigger). Differently from FlashCam, NectarCAM has a linear amplification and two gains are needed to obtain the full 0.5-2000 p.e. dynamic range~\cite{linearity}.  

The sampling and digitization of the signal is performed in the NECTAr chip~\cite{nectar0}, a switched capacitor array able to perform the sampling of the signal at 1 GHz, combined with a 12-bit analogue-to-digital converter (ADC). It acts like a circular buffer which holds the data until a camera trigger occurs.  The trigger logic is performed by sending discriminated signals into a backplane and looking for a cluster in 7 neighboring pixels. The trigger are time-stamped with a custom board built on the top of a White Rabbit client~\cite{timing_paper}.

\begin{figure}
\begin{center}
     \begin{subfigure}[c]{0.4\textwidth}
         \centering
         \includegraphics[width=0.95\textwidth]{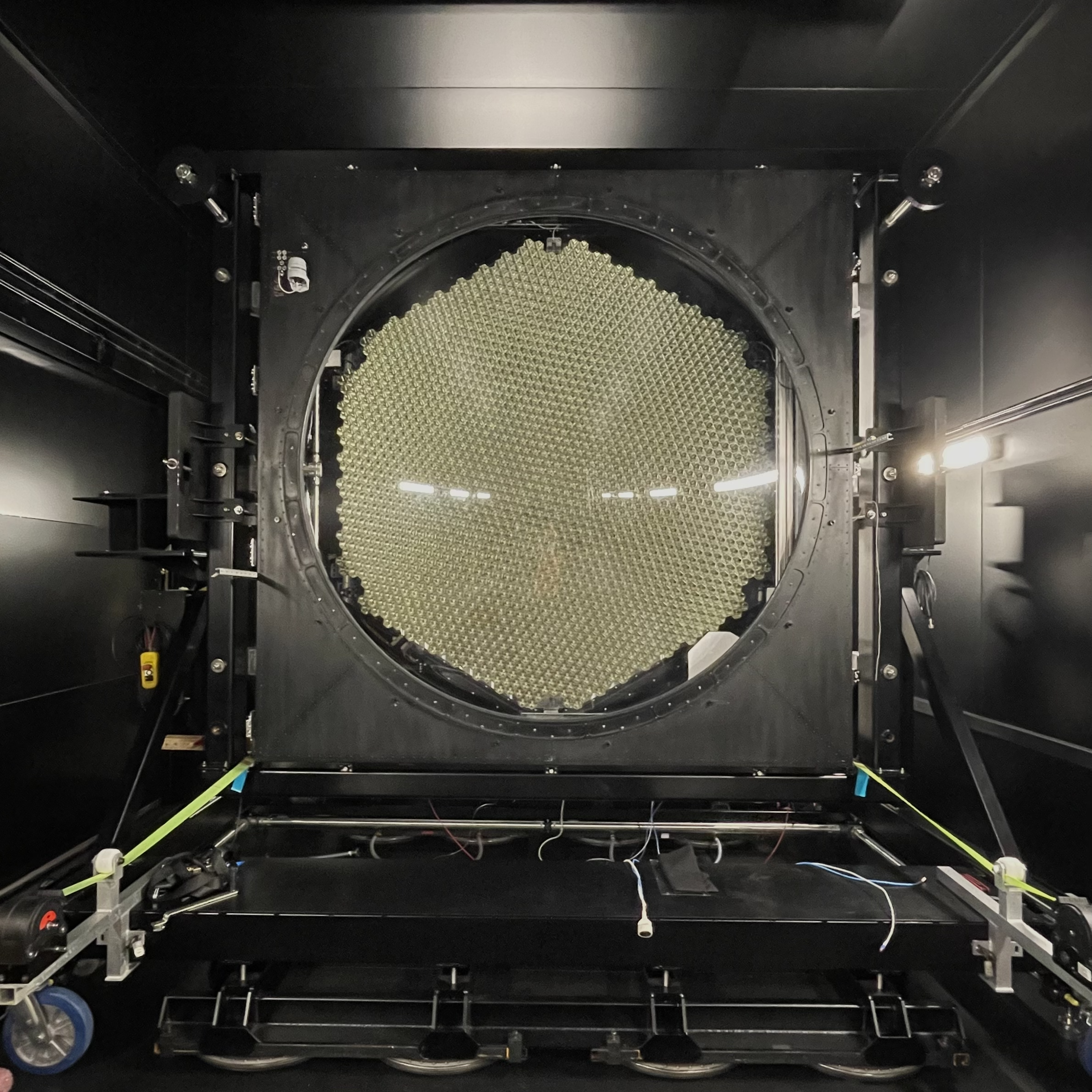}
        \caption{}
        \label{fig:nectarcam_pic}
     \end{subfigure}%
     \begin{subfigure}[]{0.6\textwidth}
         \centering
         \includegraphics[width=1\textwidth]{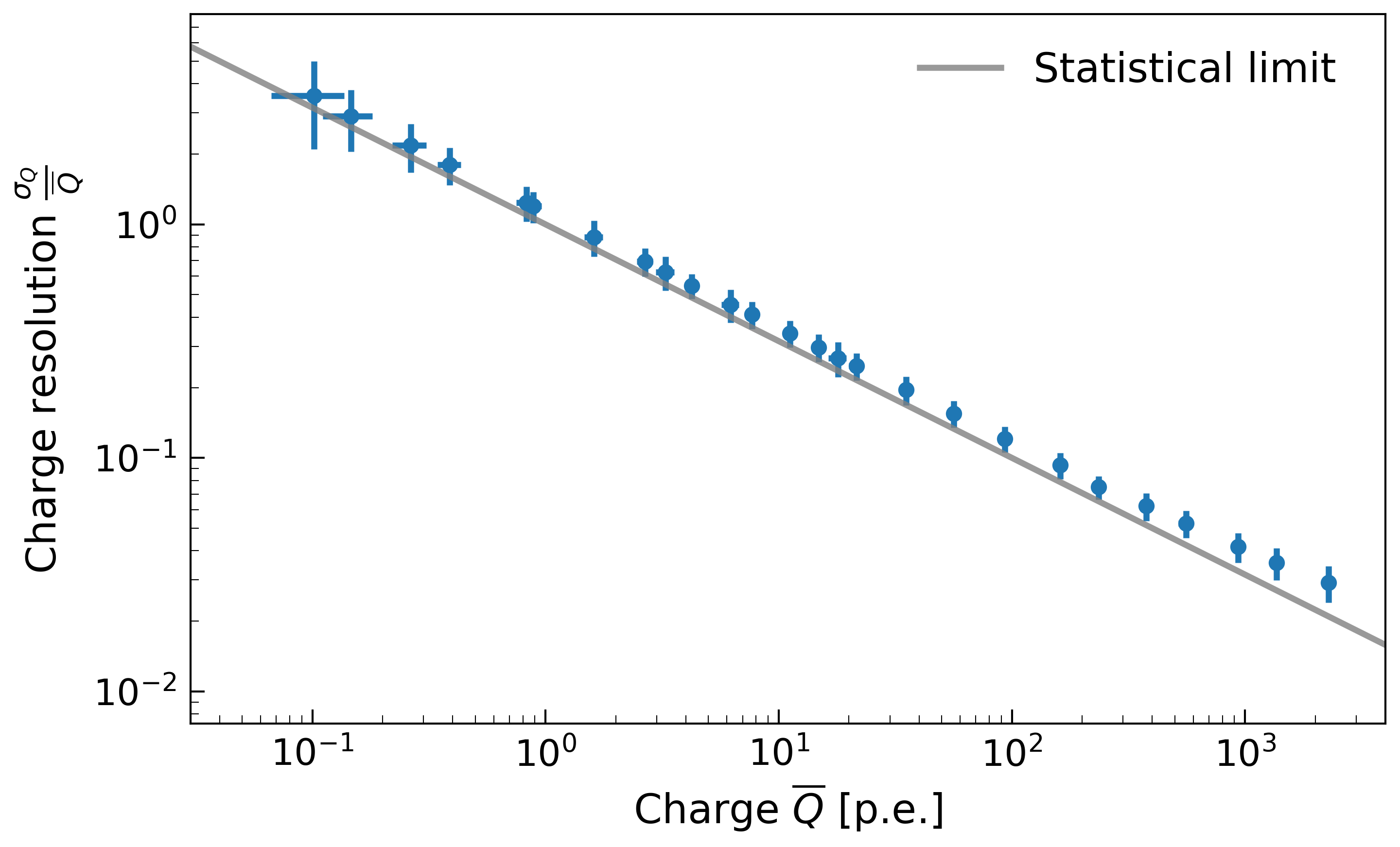}
        \caption{}
        \label{fig:nectarcam_chargeres}
     \end{subfigure}%
    \caption{\textbf{(a)}: Current image of NectarCAM at IRFU, CEA Paris-Saclay. \textbf{(b)}: Charge resolution of 10 new FEBs as a function of mean charge in p.e. measured with a LED source and filters. The statistical Poisson lower limit is shown by the gray line.}
    \label{fig:nectarcam}
    \end{center}
\end{figure}

A partially equipped camera was mounted on the MST structure prototype and successfully captured its first light in May 2019.  \autoref{fig:nectarcam_pic} illustrates the current status of the first NectarCAM, which is currently undergoing extensive verification inside a darkroom at IRFU, CEA Paris-Saclay (France). A new version of the FEB has been developed, utilizing an improved version of the NECTAr chip, which is currently undergoing comprehensive testing. The integration of this advanced NECTAr chip has significantly reduced the mimum deadtime of NectarCAM from 7~$\mu$s to 0.7~$\mu$s \cite{icrc2023_febv6}. One of the tests conducted involved measuring the charge resolution of the new FEBs, the results of which are displayed in \autoref{fig:nectarcam_chargeres}. The charge resolution was obtained by dividing the standard deviation by the mean charge in each pixel. The results align with the expected Poissonian statistical limits, and the dynamic range covers 4 decades. Further details regarding the NectarCAM camera test results can be found in \cite{timing_paper, icrc2023_spe, icrc2023_febv6, icrc2023_timing}. The camera's finalization is underway, and it is scheduled to be shipped to the Pathfinder site of CTAN in 2024.

\section{Conclusions}
We presented an overview of the current status of the MST telescope for CTAO. The MSTs are specifically designed to observe gamma-ray sources within the energy range of 100~GeV to 5~TeV. They are optimized to efficiently study extended sources, as well as investigate gamma-ray bursts and active galactic nuclei. The MST telescopes will be equipped with two different cameras: FlashCam, which has been successfully collecting data since 2019 on the HESS-CT5 telescope, and NectarCAM, which was installed on the first MST telescope prototype in Adlersolf-Berlin in the same year. Currently, both the MST structure and cameras are undergoing production and testing processes. The aim is to have them ready for installation in 2025 for the CTAO Pathfinder project, where one structure will be deployed on each site. This approach allows for on-site experience with pre-production components, enabling cost and risk reduction before initiating serial production.

\section*{Acknowledgments}
This work was conducted in the context of the CTA Consortium. We gratefully acknowledge financial support from the agencies and organizations listed here: \url{https://www.cta-observatory.org/consortium_acknowledgments/}.

\bibliographystyle{ICRC}
\bibliography{references}

%

\clearpage



\end{document}